\documentclass[aps,twocolumn]{revtex4}
\usepackage{amsmath}
\usepackage{graphicx}

\newcommand{\eq}{\begin{equation}}
\newcommand{\fine}{\end{equation}}

\begin{document}
\title{Implementation of optimal phase-covariant cloning machines}
\author{Fabio Sciarrino$^{1,2}$,\ and Francesco De Martini$^{2}$}
\address{$^{1}$Centro di\ Studi e Ricerche ''Enrico Fermi'', Via Panisperna 89/A,\\
Compendio del Viminale, Roma 00184, Italy\\
$^{2}$Dipartimento di Fisica and Consorzio Nazionale Interuniversitario per\\
le Scienze Fisiche della \ Materia, Universit\'{a} ''La Sapienza'', Roma\\
00185, Italy}

\begin{abstract}
The optimal phase covariant cloning machine (PQCM) broadcasts the
information associated to an input qubit into a multi-qubit systems,
exploiting a partial a-priori knowledge of the input state. This
additional a priori information leads to a higher fidelity than for
the universal cloning. The present article first analyzes different
experimental schemes to implement the $1\longrightarrow 3\ PQCM$.
The method is then generalized to any $1\ \longrightarrow $ $M$
machine for odd value of $M$ by a theoretical approach based on the
general angular momentum formalism. Finally different experimental
schemes based either on linear or non-linear methods and valid for
single photon polarization encoded qubits are discussed.
\end{abstract}

 \maketitle

The problem of manipulating and controlling the flux of quantum
information between many quantum systems has in general been tackled
and solved by the theory of quantum cloning and broadcasting
\cite{Scar05,Cerf05,DeMa05}. From a practical point of view, this
feature renders the theory of cloning a fundamental tool for the
analysis of the security of quantum cryptographic protocols, for the
distribution of quantum information to many partners and for the
transmission of information contained in a system into correlations
between many systems. In spite of the fact that, for fundamental
reasons, the quantum cloning and flipping operations over an unknown
qubit $\left| \phi \right\rangle $ are unrealizable in their exact
forms \cite {Woot82,Bech99}, they can be optimally approximated by
the corresponding universal quantum machines, i.e. the universal
optimal quantum cloning machine (UQCM) and the universal-NOT (U-NOT)
gate \cite{Buze96}. The optimal quantum cloning machine has been
experimentally realized following different approaches: by
exploiting the process of stimulated emission \cite
{DeMa02,Lama02,Fase02}, by means of a quantum network \cite{Cumm02}
and by adopting projective operators into the symmetric subspaces of
many qubits \cite{Ricc04,Scia04,Irvi04}. The $N\rightarrow M$ UQCM
transforms $N$ input qubits in the state $\left| \phi \right\rangle
$ into $M$ output qubits, each one in the same mixed state $\rho
_{out}.$ The quality of the copies is quantified by the fidelity
parameter ${\cal F}_{univ}^{N\rightarrow
M}=\left\langle \phi \right| \rho _{out}\left| \phi \right\rangle =\frac{%
N+1+\beta }{N+2}$ with $\beta =\frac{N}{M}\leq 1.$

Not only the ''universal'' cloning of any unknown qubit is
forbidden, but also the cloning of subsets containing non orthogonal
states. This no-go theorem ensures the security of cryptographic
protocols as $BB84$ \cite{Gisi02}. Recently {\it %
state dependent, }non universal, optimal cloning machines have been
investigated where the cloner is optimal with respect to a given
ensemble \cite{Brub00}. This partial $a-priori$ knowledge of the
state allows to reach a higher fidelity than for the universal
cloning. The simplest and most relevant case is represented by the
cloning covariant
under the Abelian group $U(1)$ of phase rotations, the so called ''{\it %
phase-covariant''} cloning. There the information is encoded in the phase $%
\phi _{i}$ of the input qubit belonging to any equatorial plane $i$
of the corresponding Bloch sphere. In this context the general state
may be expressed as: $\left| \phi _{i}\right\rangle =(\left| \psi
_{i}\right\rangle +\exp (i\phi _{i})\left| \psi _{i}^{\perp
}\right\rangle )$ and $\left\{ \left| \psi _{i}\right\rangle ,\left|
\psi
_{i}^{\perp }\right\rangle \right\} $ is a convenient normalized basis, $%
\left\langle \psi _{i}\mid \psi _{i}^{\perp }\right\rangle =0$ \cite{Brub00}%
. Precisely, in the general case the $N\rightarrow M$ phase covariant
cloning map $C_{NM}$ satisfies the following covariance relation $%
C_{NM}\left( T_{\phi i}^{\otimes N}\rho _{N}T_{\phi i}^{\dagger \otimes
N}\right) $ = $T_{\phi i}^{\otimes M}C_{NM}\left( \rho _{N}\right) T_{\phi
i}^{\dagger \otimes M}$ where $T_{\phi i}$= $\exp [-\frac{i}{2}\phi
_{i}\sigma _{i}].$ There the $\sigma _{i}$\ Pauli operator identifies the
set of input states which are cloned, e.g. $\sigma _{Y}$ corresponding to
states belonging to the $x-z$ plane of the Bloch sphere. The values of the
optimal fidelities ${\cal F}_{cov}^{N\rightarrow M}$ for this machine have
been found \cite{DAri03}. Restricting the analysis to a single input qubit
to be cloned $N=1$ into $M>1$ copies, as we do in the present paper, the
''cloning fidelity'' is found: ${\cal F}_{cov}^{1\rightarrow M}=\frac{1}{2%
}\left( 1+\frac{M+1}{2M}\right) $ for $M$ assuming odd values, or ${\cal F}%
_{cov}^{1\rightarrow M}=\frac{1}{2}\left( 1+\frac{\sqrt{M\left( M+2\right) }%
}{2M}\right) \;$for $M$ even$.$ In\ particular we have ${\cal F}%
_{cov}^{1\rightarrow 2}=0.854$ and ${\cal F}_{cov}^{1\rightarrow 3}=0.833$\
to be compared with the corresponding figures valid for universal cloning: $%
{\cal F}_{univ}^{1\rightarrow 2}=0.833$ and ${\cal F}_{univ}^{1\rightarrow
3}=0.778.$

In the above perspective it is worthwhile to enlighten the deep connection
between the cloning processes and the theory of quantum measurement \cite
{Brus98}. Indeed the concept of universal quantum cloning is related to the
problem of optimal quantum state estimation since, for $M\rightarrow \infty $%
, and $\beta \longrightarrow 0\;$the cloning fidelity converges toward the
fidelity of state estimation of an arbitrary unknown qubit: ${\cal F}%
_{univ}^{N\rightarrow M}\rightarrow {\cal F}_{estim}^{N}=\frac{N+1}{N+2}\;$%
\cite{Mass95}. In a similar way, the phase-covariant cloning is
connected with the estimation of an equatorial qubit, that is, with
the problem to find the optimal strategy to estimate the value of
the phase $\phi $ \cite {Hole82, Derk98}. The optimal strategy has
been found in \cite {Derk98}: it consists of a POVM measurement
corresponding to a von Neuman measurement onto the $N$ input qubits
characterized by a set of $N+1$ orthogonal projectors and achieves a
fidelity ${\cal F}_{phase}^{N}$. In general for $M\rightarrow \infty
,$ ${\cal F}_{cov}^{N\rightarrow
M}\rightarrow {\cal F}_{phase}^{N}.$ In particular we have ${\cal F}%
_{cov}^{1\rightarrow M}={\cal F}_{phase}^{1}+\frac{1}{4M}$ with ${\cal F}%
_{phase}^{1}=3/4.$

Recently the experimental realization of the $1\rightarrow 3$ PQCM
has been reported by adopting the methods of quantum optics
\cite{Scia05}. The present article introduces in Section I different
alternative approaches to implement the $1\rightarrow 3$ device
within any quantum information technique. In Section II such methods
are generalized to any $1\rightarrow M$ PQCM machine for odd value
of $M$. There the corresponding theoretical analysis based by on the
well established $\left| J,J_{z}\right\rangle $ angular momentum
formalism of a general $J$-spin system will be given. Finally, in
Section III different experimental schemes that can be adopted for
single photon polarization encoded qubit based either on linear and
non-linear methods will be presented.

\section{Realization of the 1$\rightarrow $3 phase-covariant cloning machine}

In the present Section we describe two different techniques to implement the
$1\rightarrow 3$ PQCM. ({\bf a}) The first method combines the
implementation of a $1\rightarrow 2$ UQCM, together with a spin flipper $%
\sigma _{i}$ and the projection of the output qubits over the symmetric
subspace: Fig. 1-({\bf a}). ({\bf b}) The second one exploits the
symmetrization of the input qubit to clone with an ancillary entangled pair:
Fig. 1-({\bf b}).

\begin{figure}[t]
\includegraphics[scale=.4]{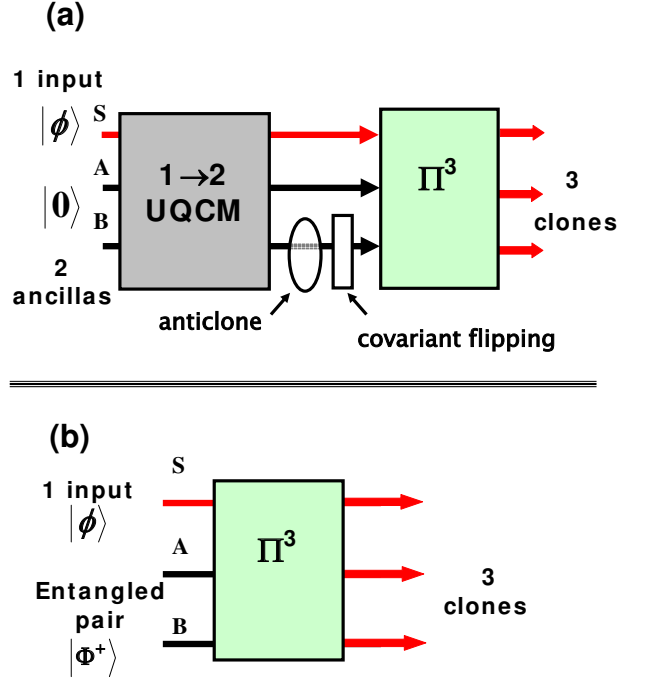} \caption{Scheme for the realization of the $1\rightarrow 3$ PQCM. ({\bf a})
UQCM, phase-covariant cloning and projection of the output state
over the symmetric subspace $\Pi ^{3}$. ({\bf b}) Symmetrization
process acting on the input qubit and one entangled pair of qubits.}
\label{fig1}
\end{figure}

We describe the approach ({\bf a}), first introduced in
\cite{Scia05}. The input qubit is expressed as: $\left| \phi
\right\rangle _{S}=2^{-%
{\frac12}%
}(\left| R\right\rangle _{S}+\exp (i\phi _{Y})\left| L\right\rangle
_{S})=\alpha \left| 0\right\rangle _{S}+\beta \left| 1\right\rangle
_{S}$, with $\left\langle R\mid L\right\rangle =0$, $\left| \alpha
\right| ^{2}+\left| \beta \right| ^{2}=1$ and $\alpha $, $\beta $
real parameters.
Here we consider, in particular, the $\phi _{Y}$ -covariant cloning and $%
\sigma _{i}=\sigma _{Y}$ realizes the NOT gate for the qubits belonging to
the $x-z$ plane. The output state of the $1\rightarrow 2$ UQCM device reads:
\begin{equation}
\begin{aligned}
&\left| \Sigma \right\rangle _{SAB}=\sqrt{\frac{2}{3}}\left| \phi
\right\rangle _{S}\left| \phi \right\rangle _{A}\left| \phi ^{\perp
}\right\rangle _{B}\\&-\frac{1}{\sqrt{6}}\left( \left| \phi
\right\rangle _{S}\left| \phi ^{\perp }\right\rangle _{A}+\left|
\phi ^{\perp }\right\rangle _{S}\left| \phi \right\rangle
_{A}\right) \left| \phi \right\rangle _{B}
\end{aligned}
\end{equation}
where the qubits $S$ and $A$ are the optimal cloned qubits while the qubit $%
B $ is the optimally flipped one. According to the scheme
represented by Fig. 1-({\bf a}), the idea is now to exactly flip the
qubit $B$ for a given subset of the Bloch sphere.
 This local
flipping transformation of $\left| \phi \right\rangle _{B}$\
leads to: $\left| \Upsilon \right\rangle _{SAB}=( I_{S}  I%
_{A}\otimes \sigma _{Y})\left| \Sigma \right\rangle _{SAB}=\sqrt{\frac{2}{3}}%
\left| \phi \right\rangle _{S}\left| \phi \right\rangle _{A}\left|
\phi \right\rangle _{B}+\\-\frac{1}{\sqrt{6}}\left( \left| \phi
\right\rangle _{S}\left| \phi ^{\perp }\right\rangle _{A}+\left|
\phi ^{\perp }\right\rangle _{S}\left| \phi \right\rangle
_{A}\right) \left| \phi ^{\perp
}\right\rangle _{B}$. By this non-universal cloning process three {\it %
asymmetric} copies have been obtained: two clones (qubits $S$ and
$A)$ with fidelity $5/6$, and a third one (qubit $B$) with fidelity
$2/3$. We may now project $S,$ $A$ and $B$ over the symmetric
subspace and obtain three symmetric clones with a higher average
fidelity. The symmetrization operator $\Pi _{SAB}^{3}$ reads as $\Pi
_{SAB}^{3}$= $\left| \Pi _{1}\right\rangle \left\langle \Pi
_{1}\right| +\left| \Pi _{2}\right\rangle \left\langle \Pi
_{2}\right| +\left| \Pi _{3}\right\rangle \left\langle \Pi
_{3}\right| +\left| \Pi _{4}\right\rangle \left\langle \Pi
_{4}\right| $ where $\left| \Pi _{1}\right\rangle =\left| \phi
\right\rangle _{S}\left| \phi \right\rangle _{A}\left| \phi
\right\rangle _{B}$, $\left| \Pi _{2}\right\rangle =\left| \phi
^{\perp }\right\rangle _{S}\left| \phi ^{\perp }\right\rangle
_{A}\left| \phi ^{\perp }\right\rangle _{B}$, $\left| \Pi
_{3}\right\rangle =\frac{1}{\sqrt{3}}\left( \left| \phi
\right\rangle \left| \phi ^{\perp }\right\rangle \left| \phi ^{\perp
}\right\rangle +\left| \phi ^{\perp }\right\rangle \left| \phi
\right\rangle \left| \phi ^{\perp }\right\rangle +\left| \phi
^{\perp }\right\rangle \left| \phi ^{\perp }\right\rangle\left|
\phi \right\rangle \right) $ and $\left| \Pi _{4}\right\rangle $= $\frac{%
1}{\sqrt{3}}\left( \left| \phi \right\rangle \left| \phi
\right\rangle \left| \phi ^{\perp }\right\rangle +\left| \phi
^{\perp }\right\rangle \left| \phi \right\rangle \left| \phi
\right\rangle +\left| \phi \right\rangle \left| \phi ^{\perp
}\right\rangle \left| \phi \right\rangle \right) $. The symmetric
subspace has dimension 4 since three qubits are involved. The
probability of success of
the projection is equal to $\frac{8}{9}$. The normalized output state $%
\left| \xi \right\rangle _{SAB}=\Pi _{SAB}^{3}\left| \Upsilon \right\rangle
_{SAB}$ is
\begin{widetext}
\begin{equation}
\left| \xi \right\rangle _{SAB}= \frac{1}{2} \sqrt{3}[\left| \phi
\right\rangle _{S}\left| \phi \right\rangle _{A}\left| \phi
\right\rangle _{B}-3^{-1}\left( \left| \phi \right\rangle _{S}\left|
\phi ^{\perp }\right\rangle _{A}\left| \phi ^{\perp }\right\rangle
_{B}+\left| \phi ^{\perp }\right\rangle _{S}\left| \phi
\right\rangle _{A}\left| \phi ^{\perp }\right\rangle _{B}+\left|
\phi ^{\perp }\right\rangle _{S}\left| \phi ^{\perp }\right\rangle
_{A}\left| \phi \right\rangle _{B}\right) ]  \label{outputPQCM}
\end{equation}
\end{widetext}
Let us now estimate the output reduced density matrices of the qubits $S,$ $%
A $ and $B:\;\rho _{S}=\rho _{A}=\rho _{B}=\frac{5}{6}\left| \phi
\right\rangle \left\langle \phi \right| +\frac{1}{6}\left| \phi ^{\perp
}\right\rangle \left\langle \phi ^{\perp }\right| $. This leads to the
fidelity ${\cal F}_{cov}^{1\rightarrow 3}=5/6$ equal to the optimal one
obtained in the general case \cite{Brub00,DAri03}. By applying a different
unitary operator $\sigma _{i}$ to the qubit $B$ we can implement the
phase-covariant cloning for the corresponding different equatorial planes of
the Bloch sphere, orthogonal to the $i-$axis.

Let us now consider the second approach ({\bf b}), which represents an
innovative simplification of the previous scheme. The PQCM device can be
realized by applying the symmetrization projection $\Pi _{SAB}^{3}$ to the
input qubit and to an ancillary entangled pair $\left| \Phi
^{+}\right\rangle _{AB}=\frac{1}{\sqrt{2}}\left( \left| 0\right\rangle
_{A}\left| 0\right\rangle _{B}+\left| 1\right\rangle _{A}\left|
1\right\rangle _{B}\right) .$ The output state reads:
\begin{equation}
\Pi _{SAB}^{3}\left( \left| \phi \right\rangle _{S}\otimes \left| \Phi
^{+}\right\rangle _{AB}\right) =\left| \xi \right\rangle _{SAB}
\end{equation}
Again the qubits $S,$ $A$ and $B$ are found to be the optimal
phase-covariant clones of the input one. By modifying the ancillary
entangled state, the set of states cloned is changed. The state
$\left| \Psi ^{+}\right\rangle _{AB}$ leads to the PQCM\ machine for
the $y-z$ plane, while $\left| \Phi ^{-}\right\rangle _{AB}$ for the
$x-y$ plane. Such result is at variance with the one found for the
universal cloning process \cite {Scia04}. Indeed the $1\rightarrow
3$ UQCM transformation can be achieved by applying the projector
$\Pi _{SAB}^{3}$ to the qubit $\left| \phi \right\rangle _{S}$ and
to two ancillas qubit, each one in a fully mixed state
$\frac{I}{2}$.

\section{General approach: 1$\rightarrow $M device}

In the present Section the two previous approaches are generalized to the
realization of the $1\rightarrow M=2P-1$ PQCM: the first one ({\bf a})
exploits the universal cloning machine, covariant flipping and final
symmetrization while the second one ({\bf b})\ is based on appropriate
symmetrization of the input qubit with entangled pairs of qubit.

\begin{figure}[h]
\includegraphics[scale=.4]{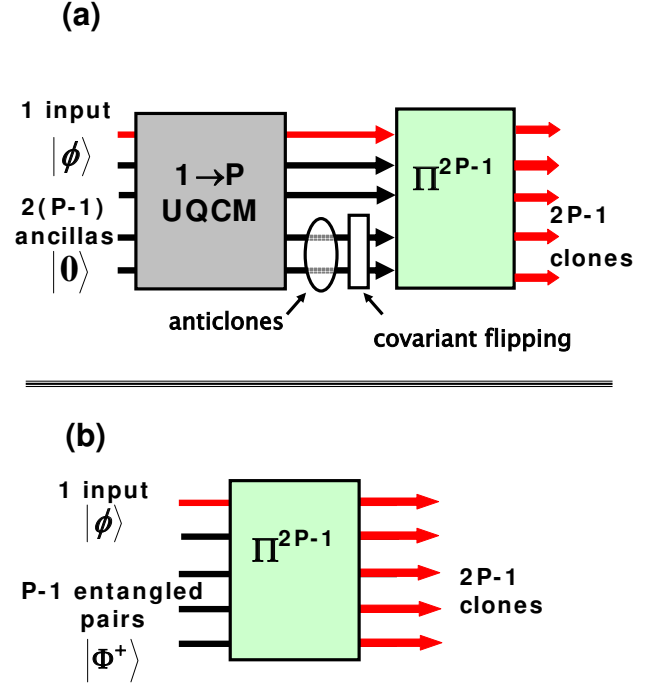} \caption{General scheme for the realization of the $1\rightarrow (2P-1)$
PQCM. ({\bf a}) $1\rightarrow P$ UQCM, phase-covariant cloning and
projection of the output state over the symmetric subspace $\Pi ^{2P-1}$. (%
{\bf b}) Symmetrization process acting on the input qubit and
$(P-1)$ entangled pairs of qubits.} \label{fig1}
\end{figure}

Let us consider the scheme of Fig.2-({\bf a}). The UQCM\ broadcast
the information on the input qubit over $2P-1$ qubit.\ The overall
output state after the UQCM map reads
\begin{equation}
\left| \Omega ^{\prime }\right\rangle =\sum_{k=0}^{P-1}b_{k}\left| \left\{
(P-k)\phi ;k\phi ^{\perp }\right\} \right\rangle _{C}\otimes \left| \left\{
k\phi ;\left( P-1-k\right) \phi ^{\perp }\right\} \right\rangle _{AC}
\end{equation}
where$\ b_{k}=\left( -1\right) ^{k}\sqrt{\frac{2}{P+1}}\sqrt{\frac{%
(P-1)!(P-k)!}{P!(P-1-k)!}}$ and the notation $\left| \left\{ p\phi ;q\phi
^{\perp }\right\} \right\rangle $ stands for a total symmetric combination
of $p$ qubits in the state $\left| \phi \right\rangle $ and of $q$ qubits in
the state $\left| \phi ^{\perp }\right\rangle $ \cite{Buze96}$.$ The labels $%
C$ and $AC$ identify, respectively, the cloning and anticloning
subsystems. Hereafter, we assume the input qubit to be in the state
$\left| \phi \right\rangle =\left| 0\right\rangle $ without lack of
generality. The $P$
qubits of the set $C$ exhibit a fidelity of the cloning process equal to $%
{\cal F}_{1\rightarrow P}=$ $\frac{2+\beta }{3}$ with $\beta =1/P,$ while
the $P-1$ qubits of the set $AC$ exhibit a fidelity of the flipping process
equal to ${\cal F}_{1\rightarrow (P-1)}^{\ast }=\frac{2}{3}$. We associate
to each qubit state a spin $\frac{1}{2}$ system. The previous expression can
hence be expressed by exploiting the formalism of the angular momentum $%
\left| J,J_{z}\right\rangle $ of a general $J-$spin system. The overall
state in the basis $\left| j;m_{j}\right\rangle _{C}\otimes \left|
j;m_{j}\right\rangle _{AC}$ reads
\begin{equation}
\left| \Omega ^{\prime }\right\rangle =\sum_{k=0}^{P-1}b_{k}\left| \frac{P}{2%
};\frac{P}{2}-k\right\rangle _{C}\otimes \left| \frac{P-1}{2};\frac{-(P-1)}{2%
}+k\right\rangle _{AC}  \label{OutputCloning}
\end{equation}
In the above representation, the overall output state of the cloner is
written as the composition of two angular momenta: ${\bf J}_{C},{\bf J}_{AC}$%
\ defined respectively over the ''cloning'' and ''anticloning'' output
channels. We note that the qubits $AC$ assume the maximum allowed value of $%
J=\frac{P-1}{2}$, thus they lie in the symmetric subspace in analogy with
the clone ones.

As following step, a covariant flipping process is applied to the subspace $%
AC$ transforming $\left| \Omega ^{\prime }\right\rangle $ into

\begin{equation}
\begin{aligned}
&\left| \Omega ^{\prime \prime }\right\rangle =I_{C}\otimes \left(
\sigma _{Y}^{\otimes (P-1)}\right) _{AC}\left| \Omega ^{\prime
}\right\rangle \\&=\sum_{k=0}^{P-1}b_{k}\left| \frac{P}{2};\frac{P}{2}%
-k\right\rangle _{C}\otimes \left| \frac{P-1}{2};\frac{(P-1)}{2}%
-k\right\rangle _{AC}
\end{aligned}
\end{equation}

Such expression holds for any qubit belonging to the equatorial
plane under consideration. Let us now express $\left| \Omega
^{\prime \prime
}\right\rangle $ adopting the overall angular momentum ${\bf J}_{T}{\bf =J}%
_{C}+{\bf J}_{AC}$ in the basis $\left|
j_{C};j_{AC};j_{T};m_{T}\right\rangle $%
\begin{equation}
\left| \Omega ^{\prime \prime }\right\rangle
=\sum_{j_{T}=1/2}^{2P-1}%
\sum_{m_{T}=-j_{T}}^{j_{T}}c(j_{T},m_{T})\left| \frac{P}{2};\frac{%
P-1}{2};j_{T};m_{T}\right\rangle
\end{equation}
where$\ c(j_{T},m_{T})$ can be derived exploiting the Clebsch -
Gordan coefficient $\left\langle j_{1};j_{2};m_{1};m_{2}\right|
\left.
j_{1};j_{2};j_{T};m_{T}\right\rangle $\ with $j_{1}=\frac{P}{2}$, $j_{2}=%
\frac{P-1}{2}$, $m_{1k}=\frac{P}{2}-k$, $m_{2k}=\frac{(P-1)}{2}-k$ \cite
{Edmonds}.

To complete the protocol, the overall output state is symmetrized by
applying the projector $\Pi ^{M}$ with $M=2P-1$ defined as: $\Pi
^{M}=\sum_{j=0}^{M}\left| \frac{P}{2};\frac{P-1}{2};\frac{M}{2};\frac{M}{2}%
-j\right\rangle \left\langle \frac{P}{2};\frac{P-1}{2};\frac{M}{2};\frac{M}{2%
}-j\right| .$ The non-vanishing contributions to the projected state
comes from terms with $j_{T}=\frac{2P-1}{2}.$ After the action of
$\Pi ^{M}$ we obtain the following normalized output state
\begin{widetext}
\begin{equation}
\left| \Omega ^{\prime \prime \prime }\right\rangle =\Pi ^{M}\left| \Omega
^{\prime \prime }\right\rangle =\sum_{k=0}^{P-1}d_{k}\left| \frac{P}{2};%
\frac{P-1}{2};\frac{2P-1}{2};\frac{2P-1}{2}-2k\right\rangle  \label{outputNM}
\end{equation}
with
\begin{eqnarray}
d_{k} &=&b_{k}\left\langle \frac{P}{2};\frac{P-1}{2};\frac{P}{2}-k;\frac{%
(P-1)}{2}-k\right| \left. \frac{P}{2};\frac{P-1}{2};\frac{2P-1}{2};\frac{2P-1%
}{2}-2k\right\rangle = \\
&=&\left( -1\right) ^{k}\sqrt{\frac{2}{P+1}}%
{P-1 \choose k}%
{2P-1 \choose 2k}%
^{-1/2}  \nonumber
\end{eqnarray}
\end{widetext}
The normalization factor reads
\begin{equation}
\left| \Pi ^{M}\left| \Omega ^{\prime \prime }\right\rangle \right| ^{2}=%
\frac{2}{P+1}\sum_{k=0}^{P-1}\frac{%
{P-1 \choose k}%
^{2}}{%
{2P-1 \choose 2k}%
}
\end{equation}

The fidelities of the phase-covariant cloning process can be inferred
re-arranging the output state (\ref{outputNM}) as follows
\begin{equation}
\left| \Omega ^{M}\right\rangle =\sum_{k=0}^{2P-1}d_{k}\left| \left\{
(2P-1-2k)\phi ;2k\phi ^{\perp }\right\} \right\rangle
\end{equation}
All the $2P-1$ qubits belonging to such state have an identical reduced
density matrix equal to
\begin{equation}
\rho _{cov}=\gamma (P)\left| \phi \right\rangle \left\langle \phi \right|
+(1-\gamma (P))\left| \phi ^{\perp }\right\rangle \left\langle \phi ^{\perp
}\right|  \label{reducedqubits}
\end{equation}
with
\[
\gamma (P)=\frac{\sum_{k=0}^{P-1}\frac{(2P-1-2k)}{(2P-1)}\frac{%
{P-1 \choose k}%
^{2}}{%
{2P-1 \choose 2k}%
}}{\sum_{k=0}^{P-1}\frac{%
{P-1 \choose k}%
^{2}}{%
{2P-1 \choose 2k}%
}}=\frac{1}{2}\left( 1+\frac{M+1}{2M}\right)
\]
The previous expression has been demonstrated numerically, for value of $M$
up to 2000.

The fidelity of the cloning process is thus
\begin{equation}
{\cal F}_{1\rightarrow M}=\left\langle \phi \right| \rho _{cov}\left| \phi
\right\rangle =\frac{1}{2}\left( 1+\frac{M+1}{2M}\right)
\end{equation}
and is found equal to the optimal one.

As alternative approach, the $1\rightarrow M$ PQCM\ device can be obtained
by applying the symmetrization projector $\Pi ^{M}$ over the input qubit and
$(P-1)$ ancilla entangled pairs $\left| \Phi ^{+}\right\rangle _{AB}$:
Fig.2-({\bf b}). Such a result can easily be obtained by manipulating the
scheme of Fig.2-({\bf a}) as follows. The UQCM of Fig. 2-({\bf a}) can be
realized starting from the input qubit $\left| \phi \right\rangle $ and $%
(P-1)$ entangled pairs $\left| \Psi ^{-}\right\rangle _{AB}$ as shown in
Ref. \cite{Scia04}. The cloning map is achieved by symmetrization of the
input qubit and $(P-1)$ ancilla qubits $A$, each one belonging to an
entangled pair $\left| \Psi ^{-}\right\rangle _{AB}$
\begin{equation}
\Pi _{SA}^{P}\otimes  I_{B}^{P-1}(\left| \phi \right\rangle
_{S}\left| \Psi ^{-}\right\rangle _{AB}^{\otimes (P-1)})
\end{equation}
The output state is equal to the one $\left| \Omega ^{\prime }\right\rangle $
of Eq.\ref{OutputCloning} up to a normalization factor. To implement the
PQCM device, the covariant flipping $\sigma _{Y}$ is then applied to the $%
(P-1)$ qubits belonging to the subset $B$. The same result can be
obtained starting from the input state $\left| \Phi
^{+}\right\rangle _{AB}^{\otimes (P-1)}$, indeed

\begin{equation}
\begin{aligned}
&\left( I_{SA}^{P}\otimes \sigma _{Y-B}^{\otimes (P-1)}\right)
\left( \Pi _{SA}^{P}\otimes I_{B}^{P-1}(\left| \phi \right\rangle
_{S}\left| \Psi ^{-}\right\rangle _{AB}^{\otimes (P-1)})\right)
\\&=\left( \Pi _{SA}^{P}\otimes I_{B}^{P-1}(\left| \phi \right\rangle
_{S}\left| \Phi ^{+}\right\rangle _{AB}^{\otimes (P-1)})\right)
\end{aligned}
\end{equation}
As final step the overall state is projected into the symmetric subspace
through the projector $\Pi _{SAB}^{2P-1}$:
\begin{eqnarray}
\left| \Omega ^{\prime \prime \prime }\right\rangle &=&\Pi
_{SAB}^{2P-1}\left( \left( \Pi _{SA}^{P}\otimes I_{B}^{P-1}(\left|
\phi \right\rangle \left| \Phi ^{+}\right\rangle _{AB}^{\otimes
(P-1)})\right) \right) = \\
&=&\Pi _{SAB}^{2P-1}\left( \left| \phi \right\rangle _{S}\left| \Phi
^{+}\right\rangle _{AB}^{\otimes (P-1)}\right)
\end{eqnarray}
In the previous expression we have exploited the concatenation
property of the symmetrization projector $\Pi _{SAB}^{2P-1}\left(
\Pi _{SA}^{P}\otimes I_{B}^{P-1}\right) =\Pi _{SAB}^{2P-1}$ which
has been demonstrated experimentally in Ref. \cite{Masu05}. This
concludes our simple proof of the scheme of Fig.2-({\bf b}).

\section{Realization by quantum optics}

In quantum optics the qubit can be implemented by exploiting the isomorphism
between the qubit state $\left| \phi \right\rangle =\alpha \left|
0\right\rangle +\beta \left| 1\right\rangle $ and the polarization state $%
\alpha \left| H\right\rangle +\beta \left| B\right\rangle $ of a
single photon. In this context it has been proposed to realize the
unitary transformation, $U_{N\rightarrow M}$, leading to the
deterministic UQCM, by means of the ''quantum injected'' optical
parametric amplification (QIOPA) in the entangled configuration. The
experimental demonstrations of both optimal cloning and flipping
processes by exploiting this technique have been reported in \
\cite{Lama02,Scia04,Irvi04}. At the same time, a different scenario
has been disclosed by the discovery that it is possible to implement
contextually the $1\rightarrow 2$ universal quantum cloning machine
(UQCM) and the $1\rightarrow 1$ universal NOT gate by modifying the
quantum state teleportation protocol \cite{Ricc04,Scia04}. The last
procedure is based on a symmetric projective operation realized by
combining single-photon interferometry and post-selection
techniques, and it can be extended to the generic $N\rightarrow M$
cloning device.

\begin{figure}[h]
\includegraphics[scale=.4]{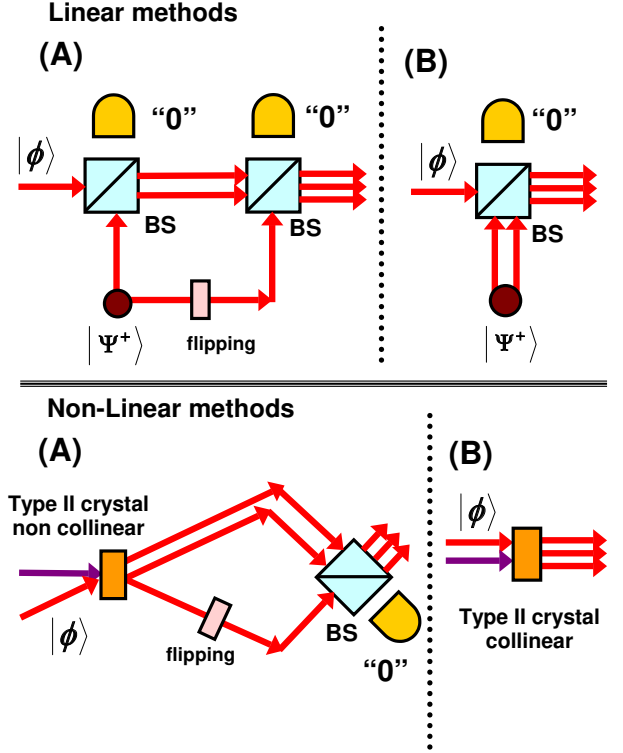} \caption{Linear methods: ({\bf a}) schematic diagram of the linear optics
multi qubit symmetrization apparatus realized by a chain of
interconnected Hong-Ou-Mandel interferometer; ({\bf b})
symmetrization of the input photon and the ancilla polarization
entangled pairs. Non-linear methods: ({\bf a}) UQCM by optical
parametric amplification, flipping by a couple of waveplates and
projection over the symmetric subspace; ({\bf b}) collinear optical
parametric amplification within a type II\ BBO crystal.}
\label{fig1}
\end{figure}

The symmetrization of two polarization encoded \ qubit can be achieved by
letting two independent-qubits to impinge onto the input arms of a beam
splitter (BS) in an Hong-Ou-Mandel interferometer \cite{Hong87}, and then by
probabilistically post-selecting the events in which the two photons emerge
in the same spatial output mode. The basic principle at the heart of these
realizations is the following: the two photons are initially superimposed at
the BS interface in order to make them indistinguishable; then, a spatial
symmetric wavefunction of the two photons is post-selected by the
measurement apparatus. Such scheme can be extended in a controlled way to an
higher number of photons, as shown in Ref.\cite{Masu05}. There a linear
optics multi qubit symmetrization apparatus has been realized by a chain of
interconnected Hong-Ou-Mandel interferometer.

Here we introduce a variety of schemes which can be realized through the
methods of quantum optics outlined above. By restricting our attention to
the $1\rightarrow 3$ PQCM, the diagram below can be easily extended to
general case $1\rightarrow M$ for odd values of $M$ following the guidelines
of the previous Section. Let us consider first linear optics approach. Fig.3
shows the experimental scheme implementing, respectively, the scheme of
Fig.1-({\bf a}), ({\bf A}), and Fig.1-({\bf b}), ({\bf B}). The flipping
operation $\sigma _{Y}$ is realized by means of two $\lambda /2$ waveplates
acting on the polarization state, while the symmetrization is implemented by
overlapping the incoming photons on a beam splitter and post-selecting the
events in which they emerge over the same mode, as said. Such scheme is
similar to the one proposed by Zou {\it et al.} \cite{Zou05} to implement
the $1\rightarrow 3$ PQCM\ for photonic qubit.

Finally the same results can be obtained adopting non-linear
methods. Let us consider the $1\rightarrow 3$ PQCM, in particular
the optimal quantum cloning for $x-z$ equatorial qubits by taking
linear polarization states as input. The $UQCM$ has been realized by
adopting a quantum-injected optical parametric amplifier (QIOPA),
while the $\sigma _{Y}$ operation and the $\Pi ^{3}$ projection have
been implemented with linear optics and post-selection
techniques Fig.({\bf a}). The flipping operation on the output mode ${\bf k}%
_{AC}$ was realized by means of two $\lambda /2$ waveplates, while the
physical implementation of the projector $\Pi ^{3}$ on the three
photons-states was carried out by linearly superimposing the modes ${\bf k}%
_{C}$ and ${\bf k}_{AC}$ on the 50:50 beamsplitter $BS$ and then by
selecting the case in which the three photons emerged from $BS$ on the same
output mode ${\bf k}_{PC}$ (or, alternatively on ${\bf k}_{PC}^{\prime }$ ).

Interestingly, the same overall state evolution can also be obtained, with
no need of the final $BS$ symmetrization, at the output of a QI-OPA with a
type II crystal working in a {\it collinear} configuration, ({\bf b}) \cite
{DeMa98}. In this case the interaction Hamiltonian $\widehat{H}_{coll}=i\chi
\hbar \left( \widehat{a}_{H}^{\dagger }\widehat{a}_{V}^{\dagger }\right)
+h.c.$ acts on a single spatial mode $k$. A fundamental physical property of
$\widehat{H}_{coll}$ consists of its rotational invariance under $U(1)$
transformations, that is, under any arbitrary rotation around the $z$-axis.
Indeed $\widehat{H}_{coll}$ can be re-expressed as $\frac{1}{2}i\chi \hbar
e^{-i\phi }\left( \widehat{a}_{\phi }^{\dagger 2}-e^{i2\phi }\widehat{a}%
_{\phi \perp }^{\dagger 2}\right) +h.c.$ for $\phi \in (0,2\pi )$ where $%
\widehat{a}_{\phi }^{\dagger }=2^{-1/2}(\widehat{a}_{H}^{\dagger }+e^{i\phi }%
\widehat{a}_{V}^{\dagger })$ and $\widehat{a}_{\phi \perp }^{\dagger
}=2^{-1/2}(-e^{-i\phi }\widehat{a}_{H}^{\dagger }+\widehat{a}_{V}^{\dagger
}) $. Let us consider an injected single photon with polarization state $%
\left| \phi \right\rangle _{in}=2^{-1/2}(\left| H\right\rangle +e^{i\phi
}\left| V\right\rangle )=\left| 1,0\right\rangle _{k}$where $\left|
m,n\right\rangle _{k}$ represents a product state with $m$ photons of the
mode $k$ with polarization $\phi $, and $n$ photons with polarization $\phi
^{\perp }$. The first contribution to the amplified state, $\sqrt{6}\left|
3,0\right\rangle _{k}-\sqrt{2}e^{i2\phi }\left| 1,2\right\rangle _{k}$ is
identical to the output state obtained with the device introduced above up
to a phase factor which does not affect the fidelity value.

\section{Conclusions}

We have introduced different schemes to implement the optimal $1\rightarrow
M>1$ phase covariant cloning machine, by exploiting either the QIOPA method
or the projection over the symmetric subspace. The introduced approaches are
probabilistic, however such feature does not spoil the main physical result
of the present procedure since the optimal fidelity value can not be
improved by any probabilistic procedure implementation \cite{Fiur04}. The
present schemes do not hold for even values of $M.$ Indeed it has been
noticed that different features affect the $1\rightarrow 2P$ and $%
1\rightarrow \left( 2P-1\right) $ PQCM maps \cite{Brub00}. Recently an
optical scheme to realize the $1\rightarrow 2$ PQCM has been proposed \cite
{Fiur03} and realized experimentally \cite{Cern06}.

The experimental realization of the different protocols with the
standard quantum optics techniques has been discussed. There we
found an answer to the question recently raised by Scarani et al.
\cite{Scar05} whether\ it is possible implement any cloning
transformation different from the universal one using amplification
through stimulated emission. We have just seen that this can it be
done directly either by linear optics elements, either by a
nonlinear, quantum injected optical parametric amplification
process. The generalization of such schemes to an higher number of
input qubits $N>1$ has been found to be non-optimal and hence
deserves further investigation.

Finally we shall enlighten that the present cloning maps are
economical, that is, do not require any extra physical resources
than the clones qubits \cite{Busc05}.

We acknowledge financial support from the Ministero della Istruzione,
dellUniversit\`{a} e della Ricerca (PRIN 2005).

\end{document}